\documentclass[12pt]{article}
\usepackage[dvips]{color}
\usepackage{epsfig}
\usepackage{amsmath}
\usepackage{graphicx}

\begin{document}
\title{\bf Particle acceleration in Horava-Lifshitz black holes}
\author{{J. Sadeghi$^{a,b}$\thanks{Email: pouriya@ipm.ir}\hspace{1mm} and
B. Pourhassan$^{a,b}$\thanks{Email: b.pourhassan@umz.ac.ir}}\\
$^a$ {\small {\em  Sciences Faculty, Department of Physics, Mazandaran University,}}\\
{\small {\em P .O .Box 47416-95447, Babolsar, Iran}}\\
$^b$ {\small {\em  Institute for Studies in Theoretical Physics and
Mathematics (IPM),}}\\
{\small {\em P.O.Box 19395-5531, Tehran, Iran}} } \maketitle
\begin{abstract}
\noindent In this paper we calculate the center-of-mass energy of
two colliding test particles near the rotating and non-rotating
Horava-Lifshitz black hole. For the case of slowly rotating KS
solution of Horava-Lifshitz black hole we compare our results with
the case of Kerr black holes. We confirm the limited value of the
center-of-mass energy for the static black holes and unlimited
value of the center-of-mass energy for the rotating black holes.
Numerically, we discuss temperature dependence of the
center-of-mass energy on the black hole horizon. We obtain the
critical angular momentum of  particles. In this limit  the
center-of-mass energy of two colliding particles in the
neighborhood of the rotating Horava-Lifshitz black hole could be
arbitrarily high. We found appropriate conditions where the
critical angular momentum could have an orbit outside the
horizon.  Finally,  we obtain center-of-mass energy corresponding to this circle orbit.
\end{abstract}
\newpage
\tableofcontents
\newpage
\section{Introduction}
Recently, Banados, Silk and West (BSW) [1] shown that free particles
falling from rest at infinity outside a Kerr black holes may collide
with arbitrarily high center-of-mass (CM) energy and hence the
maximally rotating black hole might be regarded as a
Planck-energy-scale collider. They proposed that this might lead to
signals from ultra high energy collisions such as dark matter
particles. The spinning Kerr black holes as particle accelerators
discussed by the Ref. [2], they found that the ultra-energetic
collisions cannot occur near black holes in nature. In the Ref. [3]
elastic and inelastic scattering of particles in the gravitational
field of static and rotating Kerr black holes was considered and
found that the CM energy is limited for the static and can be
unlimited for the rotating black holes. In the Ref. [4] new results
for the CM energy of particles for the non-extremal black holes are
obtained and found that the CM energy has limited value, but if one
takes into account multiple scattering the CM energy becomes
infinite [5]. These results extended to the charged spinning black
hole as Kerr-Newman [6] and non-rotating charged black holes as
Reissner-Nordstrom backgrounds [7]. In near-extremal
Reissner-Nordstrom black hole it is found that there always exists a
finite upper bound of CM energy, which decreases with the black hole
charge. From above studies we find that having infinite CM energy of
colliding particles is a generic property of a rotating black holes
[8]. This universal property
helps us to understand unknown channels of reaction between elementary particles.
Also there are similar studies in other kinds of black holes [9-15].\\
In this paper we take advantages from above studies and investigate the particle acceleration process in the Horava-Lifshitz (HL) black holes [16]. The HL
gravity is an interesting quantum gravity theory,which has stimulated an developed study on cosmology and black hole solutions [17-20]. Because it is power
counting renormalizable non-relativistic gravity theory giving up the Lorentz invariance. It is expected that the HL black hole solutions asymptotically
become Einstein gravity solutions. Significant reason for consideration of HL black hole as particle accelerator is that the slowly rotating Kerr black
hole is recovered by the slowly rotating black hole solutions in the HL gravity in the IR limit [21]. Slowly rotating black holes means that one considers
the first order of rotating parameter, it may be interpreted as a black hole arising from the breaking of spherical symmetry to axial symmetry. So, it is
interesting to calculate the CM energy of two colliding particles in the rotating HL black hole and compare it with the case of Kerr black holes. Also one
can investigate universality of unlimited CM energy on the horizon of the rotating HL black holes, which proposed in the Ref. [8].
Therefore our motivation is to examine if the BSW effect remains valid in the case of rotating black hole solutions of the modified gravity models such as HL gravity.\\
This paper organized as follows. In the section 2 we review static
HL black holes and discuss about horizon structure of several kinds
of HL black holes. In the section 3 we calculate the CM energy of
two colliding particles in the HL black holes and obtain finite
value of CM energy in the non-rotating black hole. In order to
investigate universality of CM energy we recall rotating HL black
hole in the section 4 and calculate the CM energy of the rotating HL
black holes in the section 5. Also we compare our result for slowly
rotating KS black hole with the case of Kerr black holes. Finally we
give conclusion and summarize our results.
\section{Horava-Lifshitz black holes}
In this section we recall HL black hole and discuss about horizon structure of these kind of black holes. Such discussions will be useful in the study of
particle acceleration. The four-dimensional gravity action of HL theory is given by the following expression,
\begin{equation}\label{s1}
S=\int{dt dx^{3}\sqrt{g}\bar{N}[\tilde{\mathcal{L}}_{0}+\mathcal{L}_{0}+\mathcal{L}_{1}]},
\end{equation}
where
\begin{eqnarray}\label{s2}
\tilde{\mathcal{L}}_{0}&=&\frac{2}{k^{2}}(K_{ij}K^{ij}-\lambda K^{2}) \nonumber\\
\mathcal{L}_{0}&=& -\frac{k}{2\omega^{4}}C_{ij}C^{ij}++\frac{k^{2}\mu}{2\omega^{2}}\epsilon^{ijk}R_{il}^{(3)}\nabla_{j}R_{k}^{(3)l}-\frac{k^{2}\mu^{2}}{8}R_{ij}^{(3)}R^{(3)ij}\nonumber\\
\mathcal{L}_{1}&=&\frac{k^{2}\mu^{2}}{8(1-3\lambda)}\left(\frac{1-4\lambda}{4}(R^{(3)})^{2}+\Lambda_{W}R^{(3)}-3\Lambda_{W}^{2}\right)+\mu^{4}R^{(3)},
\end{eqnarray}
where $k$, $\lambda$ and $\mu$ are constant parameters, and the Cotton tensor is defined as the following,
\begin{equation}\label{s3}
C^{ij}=\epsilon^{ijk}\nabla_{k}(R_{l}^{j}-\frac{1}{4}R\delta_{l}^{j}),
\end{equation}
also the extrinsic curvature is defined as,
\begin{equation}\label{s4}
K_{ij}=\frac{1}{2\bar{N}}(g_{ij}-\nabla_{i}N_{j}-\nabla_{j}N_{i}),
\end{equation}
Furthermore, $\bar{N}$ and $N_{i}$ are the lapse and shift functions respectively, which are used in general relativity in order to split the space-time
dimensions. Indeed, we considered the projectable version of HL gravity with detailed balanced principle, and an IR modification term $\mu^{4}R^{(3)}$.\\
As we know the vacuum metric of the HL black holes is given by [22],
\begin{equation}\label{s5}
ds^{2}=f(r)dt^{2}-\frac{dr^{2}}{f(r)}-r^{2}(d\theta^{2}+\sin^{2}\theta d\phi^{2}),
\end{equation}
where we used the natural units ($c=G=1$), and $f(r)$ is,
\begin{equation}\label{s6}
f(r)=1+(\omega-\Lambda_{W})r^{2}-\left(r[\omega(\omega-2\Lambda_{W})r^{3}+\beta]\right)^{\frac{1}{2}}.
\end{equation}
where, $\beta$ is an integration constant, $\Lambda_{W}$ and
$\omega$ are real constant parameters. There are two special cases
of the HL black holes. In the first case one consider $\beta=4\omega
M$ and $\Lambda_{W}=0$ which is called the Kehagias-Sfetsos (KS)
black hole solution [23]. In this case $f(r)$ is,
\begin{equation}\label{s7}
f_{KS}(r)=1+\omega r^{2}-\omega r^{2}\sqrt{1+\frac{4M}{\omega r^{3}}},
\end{equation}
In the second case $\beta=-\frac{\alpha^{2}}{\Lambda_{W}}$, and
$\omega=0$, which is called the Lu-Mei-Pope (LMP) black hole
solution [24]. In this case $f(r)$ is,
\begin{equation}\label{s8}
f_{LMP}(r)=1-\Lambda_{W}r^{2}-\frac{\alpha}{\sqrt{-\Lambda_{W}}}\sqrt{r}.
\end{equation}
We note here that the LMP solutions are spherically symmetric space-times, but the KS solutions are asymptotically flat space-times. In order to study
equations (7) and (8) we draw the $f(r)$ function with respect to $r$ which are shown by Fig. 1 and Fig. 2.

\begin{figure}[th]
\begin{center}
\includegraphics[scale=.4]{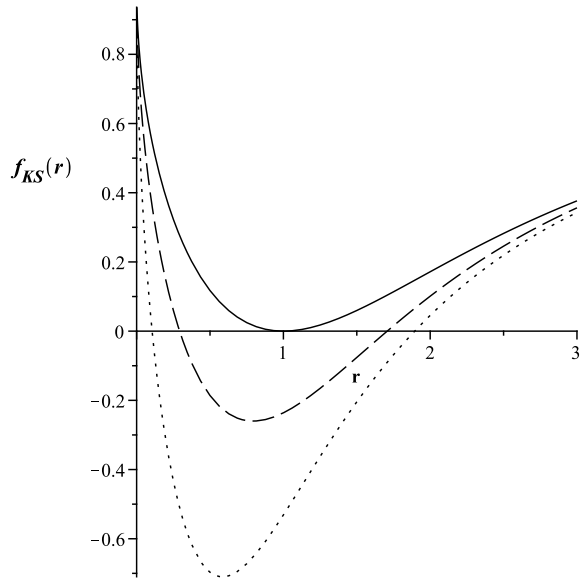}
\caption{Horizon structure of the KS solution (7) for $M=1$. The solid line represents $\omega=0.5$, where two horizons are coincide. $\omega=1$ and
$\omega=2.5$ plotted with dashed and dotted lines.}
\end{center}
\end{figure}

\begin{figure}[th]
\begin{center}
\includegraphics[scale=.4]{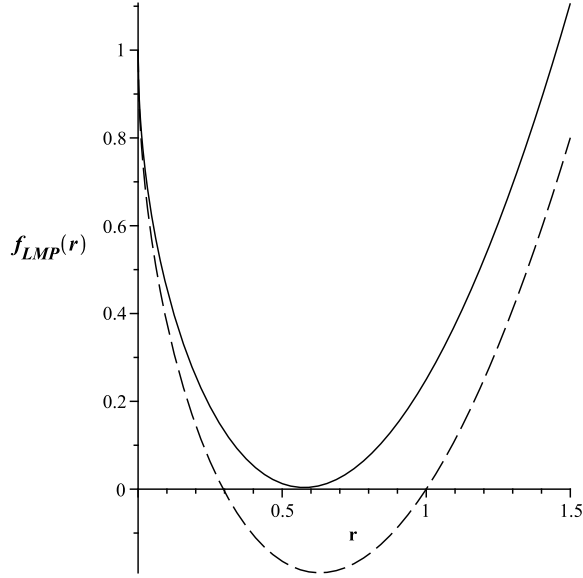}
\caption{Horizon structure of the LMP solution (8) for $\Lambda_{W}=-1$. The dashed line drawn for $\alpha=2$. And the solid line drawn for $\alpha=1.75$,
where two horizons are coincide.}
\end{center}
\end{figure}

In the KS solution (7), one can write $\omega=\frac{16\mu^{2}}{k^{2}}$, it becomes Schwarzchild black hole for $r\gg(M/\omega)^{\frac{1}{3}}$, in other
word for large $r$, or large $\omega$ ($\omega\rightarrow\infty$ limit) in fixed $r$ we have Schwarzchild black hole. The constant $\omega M^{2}$ is
dimensionless parameter and the mass $M$ is transformed to $k^{2}M$ to make mass dimension. There exist two event horizon for $\omega k^{2}M^{2}>1/2$ at,
\begin{equation}\label{s9}
r_{\pm}=k^{2}M\left(1\pm \sqrt{1-\frac{1}{2\omega k^{2}M^{2}}}\right).
\end{equation}
Two horizons of the equation (9) are coincide for $\omega k^{2}M^{2}=1/2$. We have the naked singularity for $\omega k^{2}M^{2}<1/2$. These situations
drawn in the Fig. 1 for $M=1$ and different values of $\omega$. In the Ref. [25] the motion of massless and massive test particles in the space-time of the
KS black hole solution studied. Also the Horizon structure of the LMP solution (8) drawn in the Fig. 2. By choosing $\Lambda_{W}=-1$ we find that two
horizons coincide for $\alpha=1.75$. Also we have naked singularities for the cases of
$0<\alpha<1.75$.\\
Now, it turn to back general form of the equation (6), the black hole horizon obtained by $f(r)=0$, which yields to the following equation,
\begin{equation}\label{s10}
\Lambda_{W}^{2}r^{4}+2(\omega-\Lambda_{W})r^{2}-\beta r+1=0.
\end{equation}
The equation (10) may be obtained by adding the following equations,
\begin{eqnarray}\label{s11}
\frac{1}{\Lambda_{W}^{2}r^{2}}-3r^{2}&=&\frac{2(\omega-\Lambda_{W})}{\Lambda_{W}^{2}},\nonumber\\
2r^{3}-\frac{2}{\Lambda_{W}^{2}r}&=&-\frac{\beta}{\Lambda_{W}^{2}}.
\end{eqnarray}
The first relation of (11) yields to the following answer,
\begin{equation}\label{s12}
r_{\pm}=\sqrt{\frac{\omega-\Lambda_{W}}{2\Lambda_{W}^{2}}}\left(1\pm \sqrt{1+\frac{3\Lambda_{W}^{2}}{(\omega-\Lambda_{W})^{2}}}\right).
\end{equation}
We can find that $r_{-}$ is related to the inner horizon, while $r_{+}$ is related to the outer horizon. We can also solve $\omega$ and $\beta$ from the
relations in the (11), and find,
\begin{eqnarray}\label{s13}
\omega&=&\frac{1}{2r^{2}}(1+3\Lambda_{W}r)(1-\Lambda_{W}r),\nonumber\\
\beta&=&\frac{2}{r}(1-\Lambda_{W}r^{2})(1+\Lambda_{W}r^{2}).
\end{eqnarray}
In order to have positive $\beta$ it should be
$r^{2}\leq1/\Lambda_{W}$, but in order to have positive $\omega$ it
should be $r\leq1/\Lambda_{W}$. If we assume that $\omega$ is
positive for $r=1/\Lambda_{W}$, then the upper limits of $\beta$ is
positive if $\Lambda_{W}\geq1$. On the other hand, if we assume that
$\beta$ is positive for $r^{2}=1/\Lambda_{W}$, then the upper limits
of $\omega$ is positive if $\Lambda_{W}\geq1$. It means that both
cases
constrained by the some condition $\Lambda_{W}\geq1$.\\
In the next section we calculate the CM energy of two colliding test
particles in the neighborhood of the  static HL black holes.
\section{The CM energy of two colliding particles in the static HL black holes}
In order to obtain the CM energy, we should calculate the 4-velocity
of the particles. We assume that the motion of particles is on the
equatorial plane. In that case we should set $\theta=\pi/2$ in the
equations of motion. Therefore, one can obtain,
\begin{eqnarray}\label{s14}
\dot{t}&=&-\frac{E}{f(r)},\nonumber\\
\dot{r}&=&\sqrt{f(r)\left(1+\frac{E^{2}}{f(r)}-\frac{L^{2}}{r^{2}}\right)},\nonumber\\
\dot{\theta}&=&0,\nonumber\\
\dot{\phi}&=&-\frac{L}{r^{2}},
\end{eqnarray}
where $E$ denotes the test particle energy per unit mass and $L$ denotes the angular momentum parallel to the symmetry axis per unit mass. Also the dot
denotes a derivative with respect to an affine parameter $\lambda$, which can be related to the proper time by $\tau=\eta\lambda$, and $\eta$ satisfies the
normalization condition $\eta^{2}=g_{\mu\nu}\dot{x}^{\mu}\dot{x}^{\nu}$ with $\eta=1$ for time-like geodesics and $\eta=0$ for light-like geodesics. Here,
we should note that we used sign convention $(+,-,-,-)$, therefore sign of normalization condition is different with the Ref. [15] where sign convention
used as $(-,+,+,+)$. We use nonzero 4-velocity components (14) to obtain CM energy of the two-particle collision in the background of HL black holes (5).
We assume that two particles have the angular momentum per unit mass $L_{1}$, $L_{2}$ and energy per unit mass $E_{1}$, $E_{2}$, respectively. Also we take
$m_{0}$ as the rest mass of both particles. By using the relation $E_{CM}=\sqrt{2}m_{0}\sqrt{1+g_{\mu\nu}u_{1}^{\mu}u_{2}^{\nu}}$, one can find the CM
energy of two-particle collision as the following ($u=(\dot{t}, \dot{r}, 0, \dot{\phi})$),
\begin{equation}\label{s15}
\bar{E}_{CM}=\frac{1}{f(r)r^{2}}\left(f(r)r^{2}+E_{1}E_{2}r^{2}-L_{1}L_{2}f(r)-H_{1}H_{2}\right),
\end{equation}
where,
\begin{equation}\label{s16}
H_{i}=\sqrt{f(r)r^{2}+E_{i}^{2}r^{2}-f(r)L_{i}^{2}},
\end{equation}
where $i=1,2$, and we re-scaled the CM energy as $\bar{E}_{CM}\equiv\frac{E_{CM}^{2}}{2m_{0}^{2}}$. We interest to find $\bar{E}_{CM}$ when the particles
collide on the black hole horizon. Therefore we should set $f(r)=0$ in the relation (15). It is clear that the denominator of right hand side of the
relation (15) is zero. So, if $E_{1}E_{2}>0$, then the numerator will be zero and the value of $\bar{E}_{CM}$ on the horizon will be undetermined. But when
$E_{1}E_{2}<0$, the numerator will be negative finite value and $\bar{E}_{CM}$ on the horizon will be negative infinity,
which is not physical solution.\\
In order to obtain behavior of the CM energy of two colliding particles on the horizon we expand the equation (15) at $r_{+}$. So, the lowest order term
yields to the following expression,
\begin{equation}\label{s17}
\bar{E}_{CM}(r\rightarrow r_{+})=\frac{A}{2E_{1}E_{2}r_{+}^{2}},
\end{equation}
where,
\begin{equation}\label{s18}
A\equiv r_{+}^{2}(E_{1}+E_{2})^{2}-(E_{2}L_{1}-E_{1}L_{2})^{2}.
\end{equation}
The equation (17) is just the limiting value of the relation (15) at $r\rightarrow r_{+}$. In that case we can see that if $L_{i}\rightarrow0$ or
$E_{2}L_{1}=E_{1}L_{2}$, then $\bar{E}_{CM}$ is independent of the horizon radius. We found that the value of CM energy in the horizon is finite for
$E_{i}\neq0$, as expected for non-rotating black holes.
Therefore, this case is not interesting in the present work.\\
In order to compare our result with the case of Kerr black holes and
investigate universality of having infinite energy in the center of
mass frame of colliding particles we should consider rotating HL
black holes which is subject of the next sections.
\section{Rotating HL black hole}
In this section, we consider rotating black hole in the HL gravity, which described by the following metric [21],
\begin{equation}\label{s19}
ds^{2}=f(r)dt^{2}-\frac{dr^{2}}{f(r)}-r^{2}d\theta^{2}-r^{2}\sin^{2}\theta
(d\phi-aNdt)^{2},
\end{equation}
where $a=J/M$ is the rotation parameter, and $J$ is spin angular momentum, and $M$ is the black hole mass. In order to obtain a slowly rotating black hole
solution one can keep equations of motion up to the linear order of $a$. In that case the metric (19) reduced to the following [21],
\begin{equation}\label{s20}
ds^{2}=f(r)dt^{2}-\frac{dr^{2}}{f(r)}-r^{2}d\theta^{2}-r^{2}\sin^{2}\theta
(d\phi^{2}-2aNdtd\phi).
\end{equation}
where,
\begin{equation}\label{s21}
N=\frac{2M}{r^{3}}.
\end{equation}
Also it is found that $f(r)$ in the equation (20) is KS solution which is given by the relation (7). So, the slowly rotating KS black hole solution of HL
gravity is given by,
\begin{equation}\label{s22}
ds^{2}_{slow KS}=f_{KS}(r)dt^{2}-\frac{dr^{2}}{f_{KS}(r)}-r^{2}d\theta^{2}-r^{2}\sin^{2}\theta (d\phi^{2}-\frac{4J}{r^{3}}dtd\phi).
\end{equation}
The Hawking temperature of the slowly rotating KS black hole (22) is given by the following expression,
\begin{equation}\label{s23}
T_{H}=\frac{2\omega r_{+}^{2}-1}{8\pi r_{+}(\omega r_{+}^{2}+1)},
\end{equation}
where the outer horizon $r_{+}$ is given by the equation (9). Also the angular momentum of slowly rotating black hole is given by [21],
\begin{equation}\label{s24}
J=a\left[\frac{r_{+}}{2}-\frac{3\tan^{-1}(\sqrt{\omega}r_{+})}{4\sqrt{\omega}}\right].
\end{equation}
It is found that in the $\omega\rightarrow\infty$ limit the slowly rotating KS black hole solution (22) leads to the slowly rotating Kerr solution. In the
next section we calculate CM energy of two colliding particles in the rotating HL black holes. Finally we compare results of slowly rotating KS black hole
with Kerr black holes.
\section{The CM energy of two colliding particles in the rotating HL black holes}
In this section, similar to the section 3, we calculate the CM energy of two colliding particles in the rotating HL black hole (19). In that case
4-velocity of the particles extended to the following,
\begin{eqnarray}\label{s25}
\dot{t}&=&\frac{aNL-E}{f(r)},\nonumber\\
\dot{r}&=&\sqrt{f(r)\left(1+\frac{E^{2}}{f(r)}+\frac{a^{2}N^{2}r^{2}-f(r)}{f(r)r^{2}}L^{2}-\frac{2aNEL}{f(r)}\right)},\nonumber\\
\dot{\theta}&=&0,\nonumber\\
\dot{\phi}&=&\frac{aN(aNL-E)r^{2}-f(r)L}{f(r)r^{2}}.
\end{eqnarray}
The radial component $\dot{r}$ gives us the effective potential via $2V_{eff}+\dot{r}^{2}=E^{2}$. The structure of the effective potential of the KS
solution of the static HL black hole have been discussed in the Ref. [26]. Now we present the effective potential of both KS and LMP solutions for rotating
HL black hole,
\begin{eqnarray}\label{s26}
V_{eff}(KS)&=&\frac{(L^{2}\omega-L^{2}a^{2}N^{2}-1)r^{2}-\omega
r^{4}+L^{2}+(r^{2}-L^{2})\omega r^{2}\sqrt{1+\frac{4M}{\omega
r^{3}}}}{2r^{2}},\nonumber\\
V_{eff}(LMP)&=&\frac{(L^{2}\omega-L^{2}a^{2}N^{2}-1)r^{2}-r^{4}+L^{2}+(r^{2}-L^{2})\alpha\sqrt{r}}{2r^{2}}.
\end{eqnarray}
The effective potential of the KS solution (26) in limit of $a\rightarrow0$ completely agree with the Ref. [26].\\
The structure of the effective potentials of both KS and LMP
solutions of rotating HL black hole are given in the Fig. 3. It
shows that the effective potential decreases from  $r=0$ to
$r=\infty$. As we expected the effective potential vanishes at
$r\rightarrow\infty$ and goes to infinity at $r=0$. Also the
effective potentials of the KS black hole at
$\omega\rightarrow\infty$ and $\alpha\rightarrow0$ limits goes to
the Schwarzschild black hole . From the Fig. 3 one can see that
$V_{eff}=0$ yields to the two solutions $r_{1}$ and $r_{2}$ with the
condition $r_{1}<r_{+}<r_{2}$. So, $V_{eff}(r<r_{1})$ and
$V_{eff}(r>r_{2})$ are positive, and $V_{eff}(r_{+})$ is always
negative. It means that the horizon of the HL black holes are
attractive, but $V_{eff}(r<r_{1})$ and $V_{eff}(r>r_{2})$ are
repulsive. Because of the repulsive force of the $V_{eff}(r<r_{1})$,
particles cannot fall to the center of black hole. Instead, two
particles could collide on the horizon.

\begin{figure}[th]
\begin{center}
\includegraphics[scale=.3]{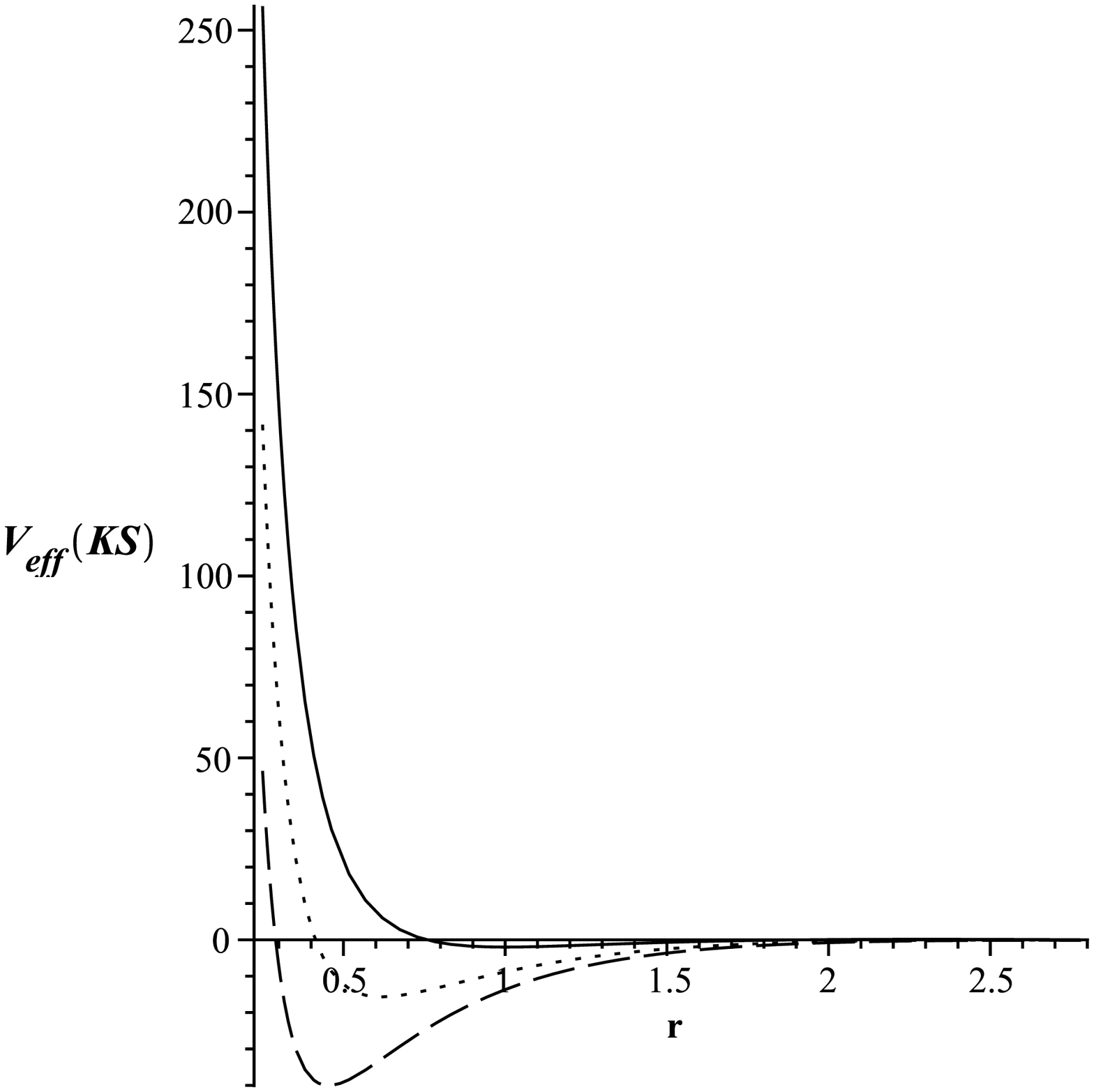}\includegraphics[scale=.3]{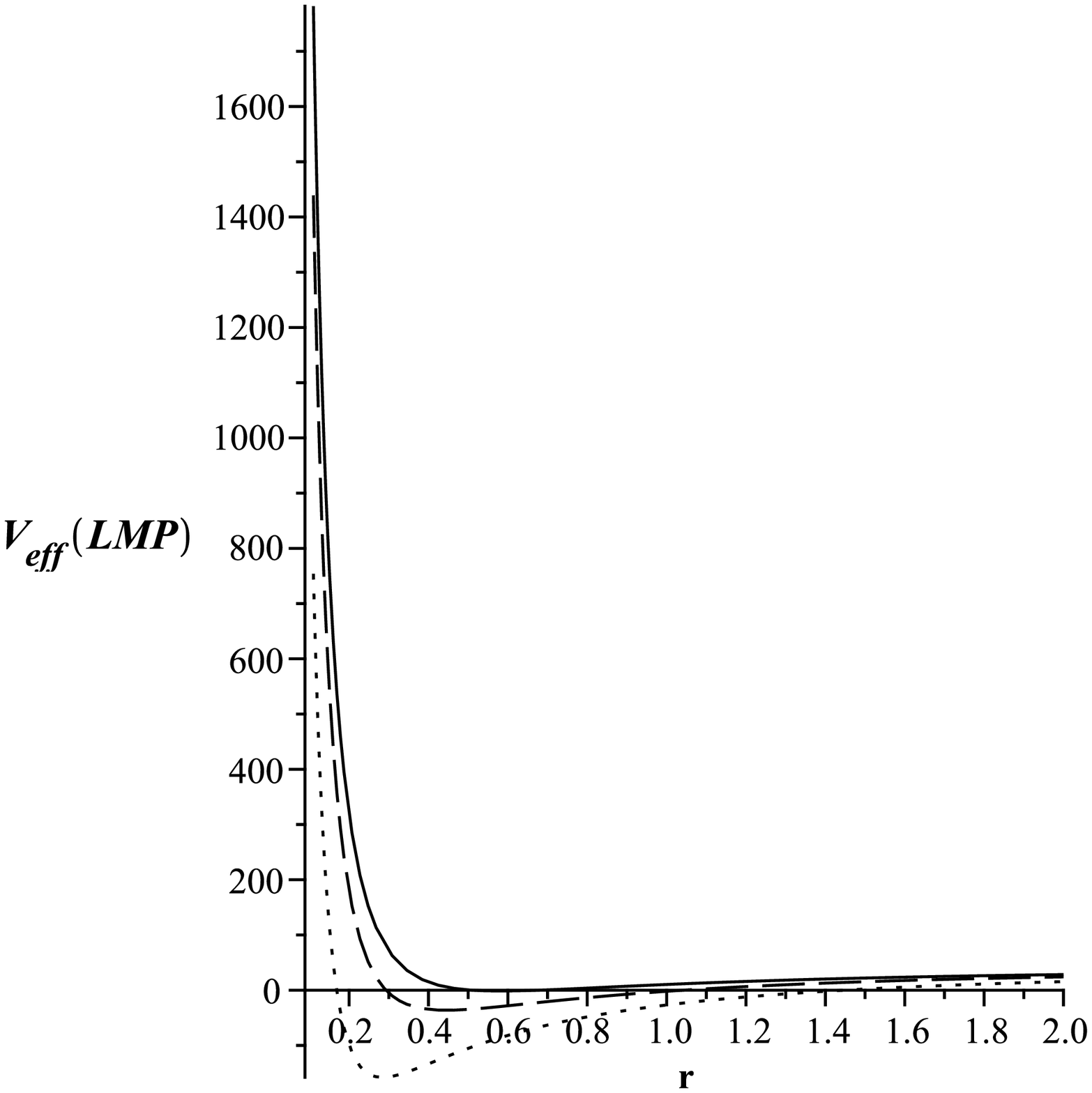}
\caption{Plots of effective potential in terms of $r$ by choosing
$M=1$, $L=10$ and $a=0.1$. Left: For KS solution, $\omega=0.5, 0.75$
and $1$ represented by solid, dotted and dashed lines respectively.
Right: For LMP solution, $\alpha=1.75, 2$ and $2.5$ represented by
solid, dashed and dotted lines respectively.}
\end{center}
\end{figure}

Then by using the 4-velocity (25) one can obtain,
\begin{equation}\label{s27}
\bar{E}_{CM}=\frac{1}{f(r)r^{2}}\left(f(r)r^{2}+E_{1}E_{2}r^{2}-L_{1}L_{2}(f(r)-a^{2}N^{2}r^{2})-aNr^{2}(E_{1}L_{2}+E_{2}L_{1})-H_{1}H_{2}\right),
\end{equation}
where,
\begin{equation}\label{s28}
H_{i}=\sqrt{f(r)r^{2}+E_{i}^{2}r^{2}-(f(r)-a^{2}N^{2}r^{2})L_{i}^{2}-2aNr^{2}E_{i}L_{i}},
\end{equation}
with $i=1,2$. The denominator of the CM energy (27) on the black hole horizon is zero, and the numerator of it reduced to
$K_{1}K_{2}-\sqrt{K_{1}^{2}}\sqrt{K_{2}^{2}}$, where,
\begin{equation}\label{s29}
K_{i}=r(aNL_{i}-E_{i}), \hspace{2cm} i=1,2,
\end{equation}
when $K_{1}K_{2}>0$, the numerator of (27) will be zero and the value of the $\bar{E}_{CM}$ on the horizon will be undetermined. On the other hand, if
$K_{1}K_{2}<0$, then $\bar{E}_{CM}$ on the horizon will be negative infinity which is not physical solution. This result agree with the result of particle
acceleration in Kerr-(anti) de Sitter black hole background [15]. It is easy to check that this result with $a=0$ reduced to the results of the section 3.
In the Fig. 4 we draw $\bar{E}_{CM}$ from the relation (27) for KS solution with small and large $\omega$ separately. We find that at large $\omega$ limit
our result agree with the Ref. [1] where the Kerr black holes living in a Minkowski space-time with a zero cosmological constant considered.

\begin{figure}[th]
\begin{center}
\includegraphics[scale=.4]{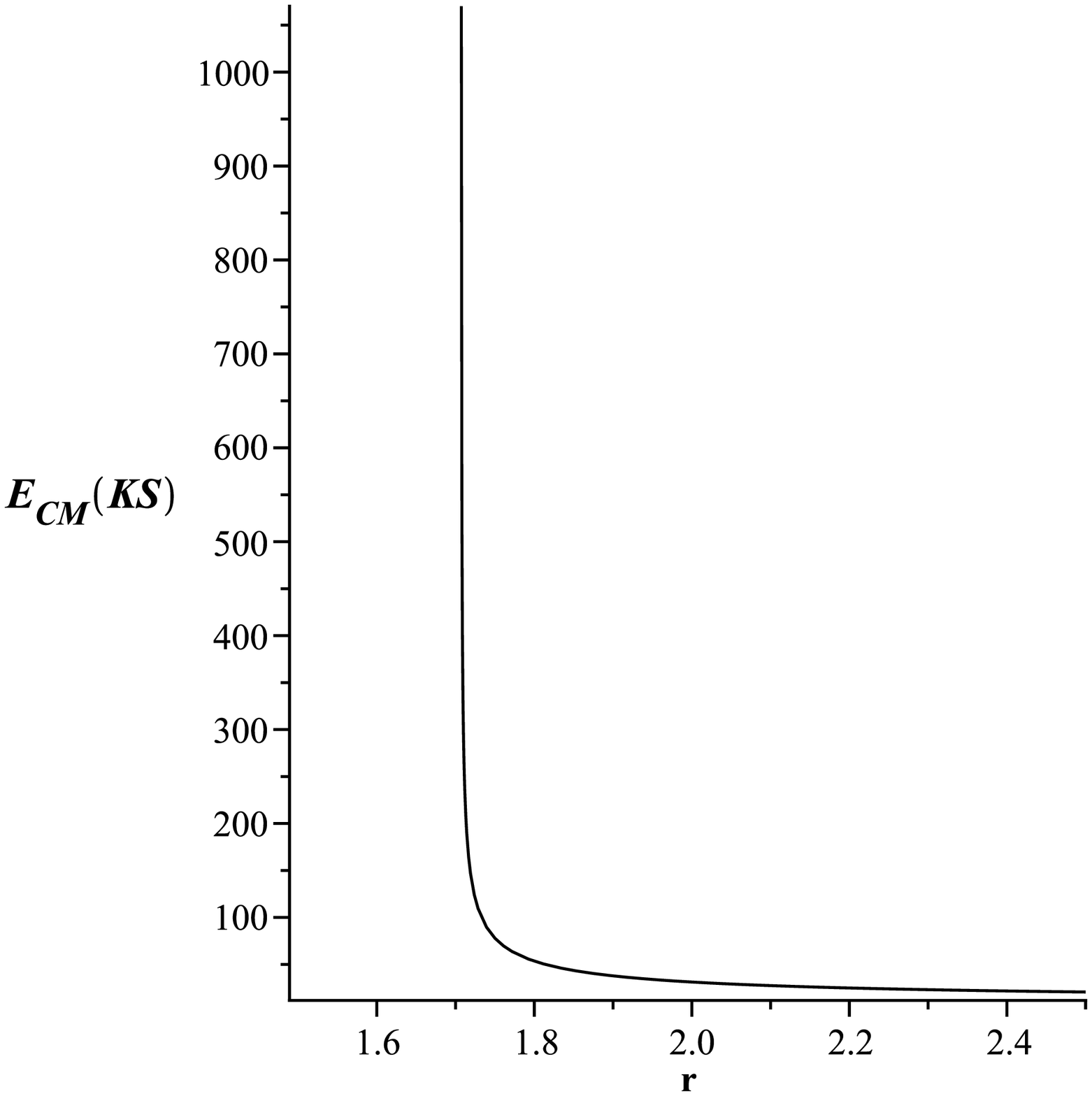}\includegraphics[scale=.4]{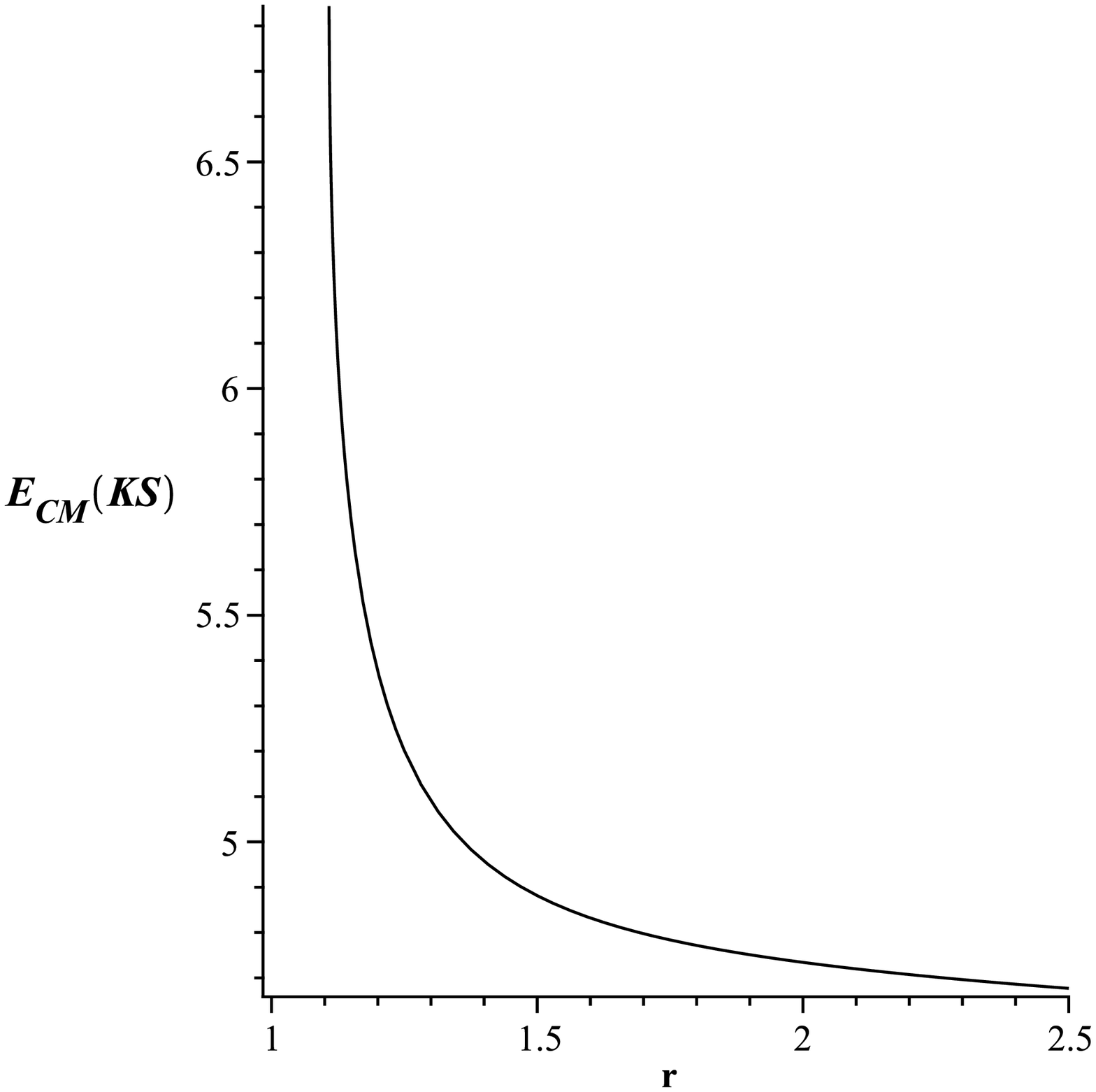}
\caption{Plots of $\bar{E}_{CM}$ in terms of $r$ by choosing $E_{1}=E_{2}=2$, $M=1$ and $L_{2}=-2$ for KS solution. Left: We take small $\omega$ and set
$L_{1}=0.5$. We can read $r_{+}\approx1.7$ from the Fig.1 . It represents $\bar{E}_{CM}$ with the critical angular momentum. Right: We take large $\omega$,
this behaves similar $\bar{E}_{CM}$ near horizon ($r_{+}=1$) of the Kerr black hole with a zero cosmological constant.}
\end{center}
\end{figure}

On the other hand the LMP black hole solution with $\Lambda_{W}=-1$
may behaves as Kerr-AdS black hole solution [22]. In the Fig. 5 we
give Plot of $\bar{E}_{CM}$ in terms of $r$ for LMP solution. It
shows that the CM energy has large value on the horizon ($r_{+}=1$).

\begin{figure}[th]
\begin{center}
\includegraphics[scale=.4]{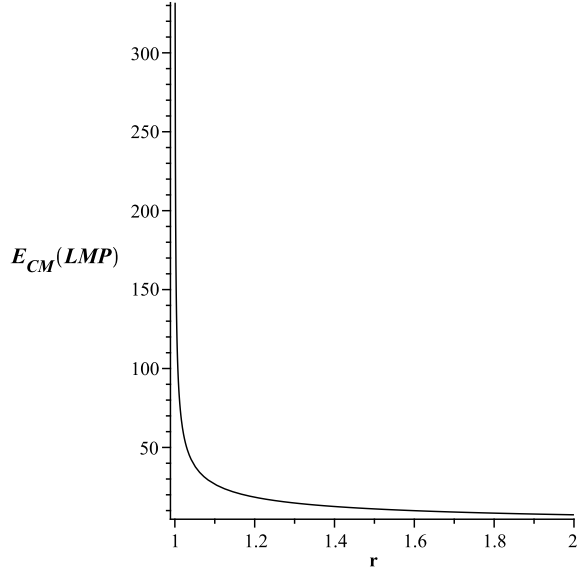}
\caption{Plot of $\bar{E}_{CM}$ in terms of $r$ by choosing
$E_{1}=E_{2}=2$, $L_{1}=0.5$, $L_{2}=-2$, $\alpha=2$ for LMP
solution. We can read $r_{+}=1$ from the Fig. 2.}
\end{center}
\end{figure}

Now, in order to obtain behavior of the $\bar{E}_{CM}$ on the
horizon we consider slowly rotating HL black hole and assume that
$a$ is small parameter, then expand $\bar{E}_{CM}$ at the horizon
and find,
\begin{equation}\label{s30}
\bar{E}_{CM}(r\rightarrow r_{+})=\frac{A}{2K_{1}(r_{+})K_{2}(r_{+})},
\end{equation}
where,
\begin{equation}\label{s31}
A\equiv (K_{1}(r_{+})+K_{2}(r_{+}))^{2}-(E_{2}L_{1}-E_{1}L_{2})^{2}.
\end{equation}
In order to study the critical angular momentum per unit mass we assume the following condition. If we consider $K_{i}(r_{+})=0$ ($aNL_{i}=E_{i}$), then
$A$ reduced to $K_{2}^{2}(1+\frac{E_{1}^{2}}{a^{2}N^{2}r_{+}^{2}})$, in that case we will obtain $(\bar{E}_{CM})|_{K_{i}=0}\rightarrow\infty$ on the
horizon. It means that, if $K_{i}(r_{+})=0$ then the CM energy of two colliding particles on the horizon of HL black hole could be arbitrary high (see Fig.
4). These lead us to obtain the critical angular momentum per unit mass from the equation (21) as the following expression,
\begin{equation}\label{s32}
L_{ci}=\frac{E_{i}}{aN}=\frac{E_{i}r_{+}^{3}}{2J}, \hspace{2cm} i=1,2,
\end{equation}
It means that the particles with the critical angular momentum $L_{ci}$ can collide with arbitrary high CM energy at the horizon. This result may provides
an effective way to probe the plank-scale physics in the background of a rotating HL black hole. We can see the critical angular momentum depends on
$r_{+}^{3}$ which is differs from the case of Kerr black hole, where the $L_{ci}$ obtained depend on $r_{+}^{2}$ [15]. However, by using the equation (24)
and taking $\omega\rightarrow\infty$ limit, we can rewrite the equation (32) as $L_{ci}\propto r_{+}^{2}$, which agree with the results of [15]. If we take
$r_{+}^{2}=-l^{2}$, we will have same critical angular momentum for two black holes ($L_{ci}(HL)=L_{ci}(Kerr)$), where $l^{2}$ is related to the
cosmological constant $\Lambda$ by $l^{-2}=-\Lambda/3$.\\
It is interesting to obtain $\bar{E}_{CM}(r_{+})$ in terms of the black hole temperature. In that case we use equations (23) and (30), and obtain the
relation between $T$ and $\bar{E}_{CM}(r_{+})$. So, numerically we find that $\bar{E}_{CM}(r_{+})$ increases with temperature. It diverges at critical
angular momentum and yields to the negative value for $L_{i}<L_{ci}$. The situation with $L_{i}>L_{ci}$ illustrated in the Fig. 6.

\begin{figure}[th]
\begin{center}
\includegraphics[scale=.4]{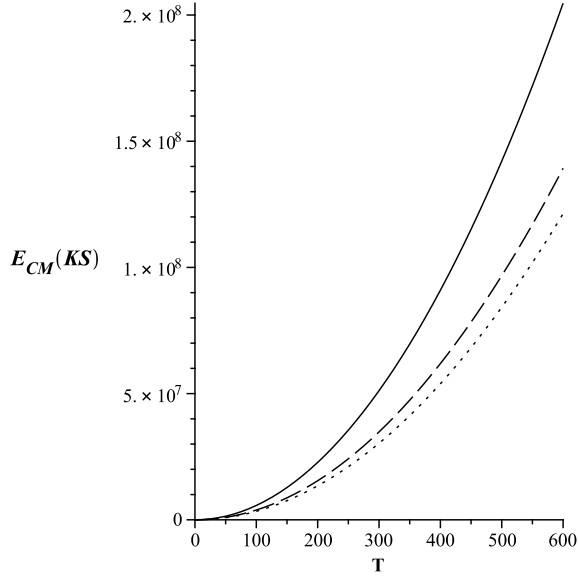}
\caption{Plot of $\bar{E}_{CM}(r_{+})$ of KS solution in terms of
the Hawking temperature for $\omega=1$, $E_{i}=2$, $L_{2}=-2$ and
$a=2$. Solid, dashed and dotted lines represent $L_{1}=1, 1.5$ and
$2$ respectively. The critical angular momentum of this case is
$L_{c1}=0.5$, where $\bar{E}_{CM}(r_{+})\rightarrow\infty$. For the
cases of $L_{c1}<0.5$ the CM energy becomes negative which is not
physical solution.}
\end{center}
\end{figure}

The particle with the critical angular momentum may have an orbit
outside the outer horizon. This will be happen if,
\begin{equation}\label{s33}
\frac{dR(r)}{dr}|_{r=r_{+}}>0,
\end{equation}
where we defined $R(r)\equiv\dot{r}^{2}$ and $\dot{r}$ is given by the relation (25). This condition comes from stability of orbital motion where
$dV_{eff}/dr$ should be positive at the balance point. By using the relation (7) and (25) one can obtain the following equation,
\begin{equation}\label{s34}
\frac{dR(r)}{dr}|_{r=r_{+}}=\frac{(a^{2}-E^{2}r_{+}^{2})W}{a^{2}},
\end{equation}
where,
\begin{equation}\label{s35}
W=2\omega r_{+}\left(1-\sqrt{1+\frac{4M}{\omega r_{+}^{3}}}\right)+\frac{6M}{r_{+}^{2}\sqrt{1+\frac{4M}{\omega r_{+}^{3}}}}.
\end{equation}
First, we assume that $a^{2}>E^{2}r_{+}^{2}$. In that case, by using the relation (9) we draw $W$ in terms of $\omega$ for KS solution of HL background. It
show that the value of $W$ is positive for $\omega>0.5$ and $M=1$ (see Fig. 7). The case of $\omega=0.5$, which yields to $W=0$, is the boundary case.
Already we found that the $\omega<0.5$ is naked singularity, therefore in the KS solution of HL black hole we have always $\omega\geq0.5$, hence there is
only $a^{2}\geq E^{2}r_{+}^{2}$ condition. $a^{2}=E^{2}r_{+}^{2}$ is the boundary case which is similar to the case of $W=0$. In summary with
$\omega\geq0.5$ and $a^{2}>E^{2}r_{+}^{2}$, the particle with the critical angular momentum can have an orbit outside the outer horizon of the HL black
hole. If this orbit be a circle, then the radial 4-velocity component of the particle must be zero, $\dot{r}=0$.\\

\begin{figure}[th]
\begin{center}
\includegraphics[scale=.4]{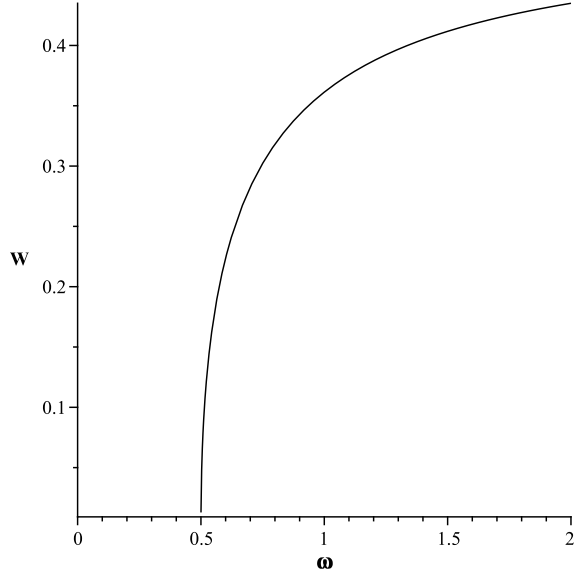}
\caption{Plot of $W$ in terms of $\omega$ for $M=1$.}
\end{center}
\end{figure}

By using the relation (25) one can obtain the angular momentum per unit mass on a circle orbit as the following,
\begin{eqnarray}\label{s36}
L_{co1}&=&\frac{aNEr^{2}+r\sqrt{f(r)}\sqrt{E^{2}+a^{2}N^{2}r^{2}-f(r)}}{a^{2}N^{2}r^{2}-f(r)},\nonumber\\
L_{co2}&=&\frac{aNEr^{2}-r\sqrt{f(r)}\sqrt{E^{2}+a^{2}N^{2}r^{2}-f(r)}}{a^{2}N^{2}r^{2}-f(r)}.
\end{eqnarray}
In order to obtain real solution we should have $E^{2}\geq
f(r)-a^{2}N^{2}r^{2}$. If $E^{2}= f(r)-a^{2}N^{2}r^{2}$, then
$L_{co1}=L_{co2}=\frac{aNr^{2}}{E}$, and there is no circle orbit.
In order to have the circle orbit, the angular momentum must be in
the range $L_{co2}<L<L_{co1}$. We assume that $L_{co1}=L_{1}$, and
$L_{co2}=L_{2}-\delta$, where $0\leq\delta\leq L_{co1}-L_{co2}$. It
means that the first particle is a target and the second particle on
the circle orbit collide with the target. $\delta$ is the small
parameter and interpreted as the drift of the second particle from
the circle orbit. In that case, one can obtain $\bar{E}_{CM}$ at the
circle orbit as the following,
\begin{equation}\label{s37}
\bar{E}_{CM}=1+\frac{E_{1}E_{2}r^{2}+Q_{1}Q_{2}}{r^{2}(a^{2}N^{2}r^{2}-f(r))}-\frac{\sqrt{f(r)}Q_{1}}{f(r)r^{2}}\delta +\mathcal{O}(\delta^{2}),
\end{equation}
where,
\begin{equation}\label{s38}
Q_{i}=r\sqrt{E_{i}^{2}+a^{2}N^{2}r^{2}-f(r)}.
\end{equation}
Here, $\delta=0$ yields to maximum of $\bar{E}_{CM}$ and
$\delta=L_{co1}-L_{co2}$ yields to minimum of $\bar{E}_{CM}$, so we
find,
\begin{eqnarray}\label{s39}
\bar{E}_{CM min}=1+\frac{E_{1}E_{2}r^{2}-Q_{1}Q_{2}}{r^{2}(a^{2}N^{2}r^{2}-f(r))},\nonumber\\
\bar{E}_{CM max}=1+\frac{E_{1}E_{2}r^{2}+Q_{1}Q_{2}}{r^{2}(a^{2}N^{2}r^{2}-f(r))},
\end{eqnarray}
and the CM energy has the rang $\bar{E}_{CM min}<\bar{E}_{CM}<\bar{E}_{CM max}$. On the other hand, if we change position of two particles
($L_{co2}=L_{1}$, and $L_{co1}=L_{2}-\delta$) then find the similar results as (39). If $f(r)=a^{2}N^{2}r^{2}$, then the maximum value of $\bar{E}_{CM}$
goes to infinity and CM energy can reach the Planck energy. For the KS solution of HL gravity, the $\bar{E}_{CM max}$ corresponding to two colliding
particles on the circle orbit with radius $r^{2}=2a$ will be infinity, where we assume that $M=1$ and $\omega$ is large. Extension of above discussions to
the case of slowly rotating is straightforward and obtained by setting $a^{2}=0$ in the relations (36) and (39).
\section{Conclusion}
Several kinds of Kerr black holes already considered as particle accelerators. In this paper,  we investigated the possibility that the Horava-Lifshitz
black holes may serve as particle accelerators. Our motivation for this study is that the slowly rotating KS solution of HL black holes reduced to slowly
rotating Kerr black holes. Therefore it is interesting to calculate CM energy of two colliding test particles in the neighborhood of the rotating HL black
holes. In agreement with the case of non-rotating Kerr black holes we obtained finite CM energy of two colliding particles on the horizon of the static HL
black holes. In this case, we found that if the angular momentum of particles vanished then the value of CM energy is independent of horizon. Then we
reviewed the rotating HL black hole and discussed that the $\omega\rightarrow\infty$ limit of the slowly rotating KS black hole solution leads to the
slowly rotating Kerr solution. Therefore,  we calculated CM energy of two colliding particles in the rotating HL black hole and found that the CM energy on
the horizon will be infinity. This result confirms the claim of the Ref. [8] where the arbitrary high CM energy of two colliding particles at the horizon
serves as general properties of rotating black holes. So, we found the critical angular momentum per unit mass where the CM energy goes to infinity. We
obtained special conditions where the particle with the critical angular momentum may have an orbit outside the outer horizon. We have shown that this
orbit may be a circle when the angular momentum limited in the special range. Hence we calculated the CM energy corresponding to this circle orbit and
found radius of circle orbit at large $\omega$. As mentioned in the section 2 we considered projectable version of HL black hole. It is also interesting to
consider non-projectable version of HL gravity [27, 28, 29, 30] and investigate BSW effect. This is currently under investigation.


\begin{thebibliography}{11}
\bibitem{P1}
M. Banados, J. Silk and S. M. West, "Kerr Black Holes as Particle
Accelerators to Arbitrarily High Energy", Phys.Rev. Lett. 103 (2009)
111102, [arXiv:0909.0169 [hep-ph]].
\bibitem{P2}
Ted Jacobson, Thomas P. Sotiriou, " Spinning Black Holes as Particle
Accelerators", Phys.Rev.Lett.104 (2010) 021101
[arXiv:0911.3363[gr-qc]]. Kayll Lake, "Particle Accelerators inside
Spinning Black Holes", Phys.Rev.Lett.104 (2010) 211102,
[arXiv:1001.5463 [gr-qc]].
\bibitem{P3}
A. A. Grib, Yu. V. Pavlov, "On Particle Collisions in the
Gravitational Field of Black Hole", Astropart. Phys.34:581-586, 2011
[arXiv:1001.0756 [gr-qc]].
\bibitem{P4}
A. A. Grib, Yu. V. Pavlov, " On Particle Collisions and Extraction
of Energy from a Rotating Black Hole", [arXiv:1004.0913 [gr-qc]].
\bibitem{P5}
A. A. Grib, Yu. V. Pavlov "On particles collisions near rotating
black holes", [arXiv:1010.2052 [gr-qc]].
\bibitem{P6}
Shao-Wen Wei, Yu-Xiao Liu, Heng Guo, Chun-E Fu, "Charged Spinning
Black Holes as Particle Accelerators", Phys.Rev.D82 (2010) 103005,
[arXiv:1006.1056 [hep-th]].
\bibitem{P7}
Oleg B. Zaslavskii, "Acceleration of particles by nonrotating
charged black holes", JETP Letters 92, 571 (2010), [arXiv:1007.4598
[gr-qc]].
\bibitem{P8}
Oleg B. Zaslavskii, "Acceleration of particles as universal property
of rotating black holes", Phys. Rev. D82 (2010) 083004,
[arXiv:1007.3678v2 [gr-qc]].
\bibitem{P9}
Oleg B. Zaslavskii, "Acceleration of particles by black holes --
general explanation", [arXiv:1011.0167 [gr-qc]].
\bibitem{P10}
Masashi Kimura, Ken-ichi Nakao, Hideyuki Tagoshi, "Acceleration of
colliding shells around a black hole -- Validity of test particle
approximation in BSW process", Phys.Rev.D83:044013,2011
[arXiv:1010.5438 [gr-qc]].
\bibitem{P11}
Maximo Banados, Babiker Hassanain, Joseph Silk, Stephen M. West,
"Emergent Flux from Particle Collisions Near a Kerr Black Hole",
Phys.Rev.D83:023004,2011 [arXiv:1010.2724 [astro-ph.CO]].
\bibitem{P12}
Tomohiro Harada, Masashi Kimura, "Collision of an innermost stable
circular orbit particle around a Kerr black hole",
Phys.Rev.D83:024002,2011 [arXiv:1010.0962 [gr-qc]].
\bibitem{P13}
Pu-Jian Mao, Ran Li, Lin-Yu Jia, Ji-Rong Ren, "Kaluza-Klein Black
Hole as Particles Accelerators", [arXiv:1008.2660 [hep-th]].
\bibitem{P14}
Shao-Wen Wei, Yu-Xiao Liu, Hai-Tao Li, Feng-Wei Chen, "Particle
Collisions on Stringy Black Hole Background", JHEP 1012 (2010) 066,
[arXiv:1007.4333 [hep-th]].
\bibitem{P15}
Yang Li, Jie Yang, Yun-Liang Li, Shao-Wen Wei, Yu-Xiao Liu,
"Particle Acceleration in Kerr-(anti-) de Sitter Black Hole
Backgrounds", [arXiv:1012.0748 [hep-th]]. Changqing Liu, Songbai
Chen, Chikun Ding, and Jiliang Jing, "Kerr-Taub-NUT black hole as
Particle Accelerators", [arxiv:1012.5126 [hep-th]].
\bibitem{P16}
Petr Horava, "Spectral Dimension of the Universe in Quantum Gravity
at a Lifshitz Point", Phys.Rev.Lett.102 (2009) 161301,
[arXiv:0902.3657 [hep-th]]. Petr Horava, "Quantum Gravity at a
Lifshitz Point", Phys.Rev.D79 (2009) 084008, [arXiv:0901.3775
[hep-th]]. Petr Horava, "Membranes at Quantum Criticality", JHEP
0903 (2009) 020, [arXiv:0812.4287 [hep-th]]. Petr Horava, "Quantum
Criticality and Yang-Mills Gauge Theory", Phys.Lett.B694 (2010) 172,
[arXiv:0811.2217 [hep-th]].
\bibitem{P17}
G. Calcagni, "Cosmology of the Lifshitz universe", JHEP 0909, 112
(2009) [arXiv:0904.0829 [hep-th]].
\bibitem{P18}
R. G. Cai, L. M. Cao and N. Ohta, "Topological Black Holes in
Horava-Lifshitz Gravity", Phys. Rev. D 80 (2009) 024003
[arXiv:0904.3670 [hep-th]]. R. G. Cai, Y. Liu and Y. W. Sun, " On
the z=4 Horava-Lifshitz Gravity", JHEP 0906, 010 (2009)
[arXiv:0904.4104 [hep-th]].
\bibitem{P19}
B. R. Majhi, "Hawking radiation and black hole spectroscopy in
Horava-Lifshitz gravity", Phys. Lett. B 686 (2010) 49
[arXiv:0911.3239 [hep-th]].
\bibitem{P20}
Y. S. Myung and Y. W. Kim, "Thermodynamics of Horava-Lifshitz black
holes", [arXiv:0905.0179 [hep-th]].
\bibitem{P21}
 Hyung Won Lee, Yong-Wan Kim, Yun Soo Myung, "Slowly rotating black holes in the Horava-Lifshitz
 gravity", Eur.Phys.J.C70:367-371,2010 [arXiv:1008.2243 [hep-th]].
\bibitem{P22}
I. Radinschi, F. Rahaman, A. Banerjee, "On the energy of
Horava-Lifshitz black holes", [arXiv:1012.0986 [gr-qc]].
\bibitem{P23}
A. Kehagias and K. Sfetsos, "The black hole and FRW geometries of non-relativistic gravity", Physics Letters B 678 (2009) 123.
\bibitem{P24}
H. Lu, Jianwei Mei, C.N. Pope, "Solutions to Horava Gravity",
Phys.Rev.Lett.103 (2009) 091301, [arXiv:0904.1595 [hep-th]].
\bibitem{P25}
V. Enolskii, B. Hartmann, V. Kagramanova, J. Kunz, C. Laemmerzahl,
P. Sirimachan, "Particle motion in Horava-Lifshitz black hole
space-times", [arXiv:1106.4913 [gr-qc]].
\bibitem{P26}
Bogeun Gwak and Bum-Hoon Lee, "Particle probe of Horava-Lifshitz gravity", JCAP09(2010)031, [arXiv:1005.2805].
\bibitem{P27}
A. Abdujabbarov, B. Ahmedov, B. Ahmedov, "Energy Extraction and
Particle Acceleration Around Rotating Black Hole in Horava-Lifshitz
Gravity", [arXiv:1107.5389 [astro-ph.SR]].
\bibitem{P28}
Elias Kiritsis, Georgios Kofinas, "On Horava-Lifshitz "Black
Holes"", [arXiv:0910.5487v1 [hep-th]].
\bibitem{P29}
G. Koutsoumbas, P. Pasipoularides, "Black hole solutions in
Horava-Lifshitz Gravity with cubic terms", [arXiv:1006.3199
[hep-th]].
\bibitem{P30}
G. Koutsoumbas, E. Papantonopoulos, P. Pasipoularides, M. Tsoukalas,
"Black Hole Solutions in 5D Horava-Lifshitz Gravity",
[arXiv:1004.2289v2 [hep-th]].
\end{thebibliography}
\end{document}